\documentclass[pre,12pt,showpacs,tightenlines]{revtex4}

\usepackage{graphicx}

\newcommand{\eq}[1]{(\ref{#1})}
\newcommand{\average}[1]{\langle #1\rangle}
\newcommand{\mathtiny}[1]{\mbox{\tiny $#1$}}
\newcommand{\ddh}[2]{\frac{d#1}{d#2}}

\begin{document}

\title{Power and Heat Fluctuation Theorems for Electric Circuits}

\author{R.\ van Zon$^1$, S.\ Ciliberto$^2$ and E.\ G.\ D.\ Cohen$^1$}

\affiliation{$^1$The Rockefeller University, 1230 York Avenue, New
York, New York 10021, USA\\$^2$Laboratoire de Physique, CNRS UMR
5672, Ecole Normale Sup\'erieure de Lyon, 46 All\'ee d'Italie, 69364
Lyon Cedex 07, France}

\date{March 31, 2004}

\begin{abstract}
Using recent fluctuation theorems from nonequilibrium statistical
mechanics, we extend the theory for voltage fluctuations in electric
circuits to power and heat fluctuations. They could be of particular
relevance for the functioning of small circuits. This is done for a
parallel resistor and capacitor with a constant current source for
which we use the analogy with a Brownian particle dragged through a
fluid by a moving harmonic potential, where circuit-specific analogues
are needed on top of the Brownian-Nyquist analogy. The results may
also hold for other circuits as another example shows.
\end{abstract}


\maketitle

Nanotechnology is quickly getting within reach, but the physics at
these scales could be different from that at the macroscopic scale. In
particular, large fluctuations will occur, with mostly unknown
consequences.  In this letter, we will investigate properties of
electric circuits concerning the fluctuations of power and heat
within the context of the so-called \emph{Fluctuation Theorems}
(FTs). These theorems were originally found in the context of
non-equilibrium dynamical systems theory. Surprisingly, these can be
applied also to electric circuits, as we will show, and thus give
further insight into their behavior.

Let us first give a brief introduction to the FTs.  First found in
dynamical systems\cite{Evansetal93,GallavottiCohen95a} and later
extended to stochastic systems\cite{Kurchan98}, these conventional FTs
give a relation between the probabilities to observe a positive value
of the (time averaged) ``entropy production rate'' and a negative
one. This relation is of the form $P(\sigma)/P(-\sigma) =
\exp[\sigma\tau]$, where $\sigma$ and $-\sigma$ are equal but opposite
values for the entropy production rate, $P(\sigma)$ and $P(-\sigma)$
give their probabilities and $\tau$ is the length of the interval over
which $\sigma$ is measured. In these systems, the above mentioned FT
is derived for a mathematical quantity $\sigma$, which has a form
similar to that of the entropy production rate in Irreversible
Thermodynamics.

Apart from an early experiment in a turbulent
flow\cite{otherexperiment}, for quite some time, the investigations of
the FTs were restricted to theoretical approaches and simulations. In
2002, Wang {\em et al.} performed an experiment on a micron-sized
Brownian particle dragged through water by a moving optical
tweezer. In this experiment, a \emph{Transient Fluctuation Theorem}
(TFT) was demonstrated for fluctuations of the total external work
done on the system in the transient state of the system, i.e.,
considering a time interval of duration $\tau$ which starts
immediately after the tweezer has been set in
motion\cite{Wangetal02}. In contrast, a \emph{Stationary State
Fluctuation Theorem} (SSFT), which was not measured, would concern
fluctuations in the stationary state, i.e., in intervals of duration
$\tau$ starting at a time long after the tweezer has been set in
motion. While the work fluctuations satisfy the conventional TFT and
SSFT\cite{MazonkaJarzynski99,VanZonCohen02b}, the heat fluctuations
satisfy different, extended FTs due to the interplay of the stochastic
motion of the fluid with the deterministic harmonic potential induced
by the optical tweezer\cite{VanZonCohen03a,VanZonCohen03b}.  Given the
possible problems with identifying the entropy
production\cite{Wangetal02,MazonkaJarzynski99}, in this letter we
prefer to consider the work and the heat.

We remark that the \emph{conventional} FTs hold in time-reversible,
chaotic dynamical systems\cite{Evansetal93,GallavottiCohen95a} and in
finite stochastic systems if transitions can occur forward and
backward\cite{Kurchan98}. However, the general condition for an
\emph{extended} FT is unknown.

In view of the well-known analogy of Brownian motion (as in the
experiment of Wang {\em et al.}) and Nyquist noise in electric
circuits\cite{VanKampenMazo}, one could ask whether the TFT and SSFT
based on the Langevin equation also apply to electric circuits.
Electric circuits are interesting as they are directly relevant to
nanotechnology and because they lend themselves easily to experiments.
Indeed, it turns out the conventional FTs hold for \emph{work} and the
extended FTs for \emph{heat}, as we will show here.  We emphasize
that to connect with previous papers on which this one is
based, we will also use the term \emph{work} for the electric
circuits, which is nothing but the time integral of the \emph{power}.

To exploit the analogy of the Langevin descriptions for electric
circuit and that of the Brownian particle, we first recall
the form of the Langevin equation for the Brownian particle in the
experiment of Wang {\em et al.}\cite{VanZonCohen02b}:
\begin{eqnarray}
  m \ddh{^2x_t}{t^2} = - \alpha \ddh{x_t}{t} + \xi_t - k (x_t -v^* t).
\label{eomBrown}
\end{eqnarray}
Here, $m$ is the mass of the Brownian particle, $x_t$ is its position
at time $t$, $\alpha$ is the (Stokes') friction coefficient, $k$ is
the strength of the harmonic potential induced by the optical tweezer,
and $v^*$ is the constant speed at which the tweezer is moved.
$\xi_t$ represents a Gaussian white noise which satisfies
\begin{equation}
  \average{\xi_t} = 0;\quad \average{\xi_t\xi_{t'}} =
  2k_BT\alpha\,\delta(t-t'),
\label{Brownnoise}
\end{equation}
where $\average{}$ denotes an average over an ensemble of similar
systems\cite{VanKampenMazo},
$T$ is the temperature
of the water and $k_B$ is Boltzmann's constant. Usually, the velocity
relaxes quickly, so one can set $m=0$ in Eq.~\eq{eomBrown}.

\begin{figure}[b]
\centerline{\includegraphics[height=0.21\textheight]{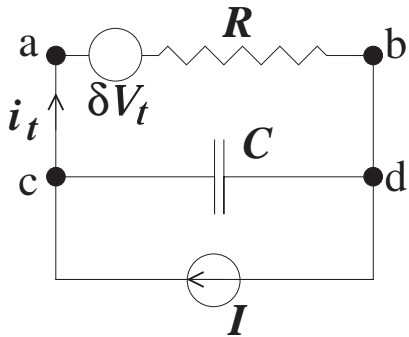}}
\caption{Circuit with Nyquist noise and a resistor and capacitor in
parallel, subject to a current source.}
\label{fig1}
\end{figure}

Next we consider in Fig.~\ref{fig1} an electric circuit in which a
resistor with resistance $R$ and a capacitor with capacitance $C$ are
arranged \emph{in parallel} and are subject to a constant,
nonfluctuating current source $I$. Energy is being dissipated in the
resistor and according to the fluctuation-dissipation theorem this
means there are fluctuations too.  To a good approximation, one
can use a Gaussian random noise term $\delta V_t$ to describe these
fluctuations\cite{VanKampenMazo}, which is depicted in the figure by a
voltage generator. In addition, we define $q_t$ as the
charge that has gone through the resistor, $i_t$ as the current that
is going through it (so $i_t=dq_t/dt$) and $q'_t$ as the charge on the
capacitor, all at time $t$.  Using that $q_t'=\int_0^tdt'\,
(I-i_{t'})=It-q_t$ and that $V_{ab}=q'_t/C$, standard calculations for
electric circuits give
\begin{eqnarray}
  0 = - R \ddh{q_t}{t} - \delta V_t - \frac{q_t -I\,t}{C},
\label{eomparallel}
\end{eqnarray}
in which $\delta V_t$ satisfies\cite{VanKampenMazo}
\begin{equation}
  \average{\delta V_t} = 0;\quad \average{\delta V_t\delta V_{t'}} =
  2k_BT R\,\delta(t-t').
\label{noise}
\end{equation}
We see that Eqs.~\eq{eomparallel} and \eq{noise} are of a very similar
form as those for a one-dimensional massless Brownian particle dragged
through a fluid by means of a harmonic potential in Eqs.~\eq{eomBrown}
and \eq{Brownnoise}\cite{footnote11}.  Table~\ref{table} gives all the
analogues, i.e., both the well-known Brownian motion-Nyquist noise
ones ($\xi_t$, $\alpha$, $T$ vs.\ $\delta V_t$, $R$, $T$) as well as
additional circuit-specific ones.

\begin{table}[t]
\centerline{
\begin{tabular}{|c|c|}
\hline
 Brownian particle & RC circuits\\\hline
  $\xi_t$&$-\delta V_t$ \\
  $\alpha$  & $R$\\
  $T$ & $T$\\
  $x_t$ & $q_t$ \\
  $v_t$ & $i_t$ \\
  $k$ & $1/C$\\
  $v^*$  & $\:I$ (or $CA$ in the serial case)\\
\hline
\end{tabular}
}
\caption{The analogy of a circuit and a
  Brownian particle.}\label{table}
\end{table}

Given this analogy, we turn to the \emph{heat} fluctuations in this
circuit.  This heat is developed in the resistor. Thus, the dissipated
heat over a time $\tau$ is given by the time integral of the voltage
over the resistor, $V_{ab}$, times the current through it, $i_t$,
i.e.,
\begin{eqnarray}
  Q_\tau =
\int_0^\tau dt\,  i_t [i_t R + \delta V_t] =
- \int_0^\tau dt\,  i_t \frac{q_t-It}{C},
\label{5}
\end{eqnarray}
where we used Eq.~\eq{eomparallel}. This is precisely the quantity
found in Refs.~\cite{VanZonCohen03a,VanZonCohen03b} for the Brownian
particle, when we use the analogies in Table~\ref{table}. Hence
$Q_\tau$ in this parallel RC circuit behaves completely analogous to
the heat for the Brownian particle and thus we know that it satisfies
the extended FT. That is, defining a \emph{fluctuation function} by
\begin{equation}
  f^{\mathtiny Q}_\tau \equiv \frac{k_BT}{\average{Q_\tau}} \ln \left[\frac{
P(+Q_\tau)}{P(-Q_\tau)}\right]
\end{equation}
and a scaled heat fluctuation by $p_{\mathtiny Q}\equiv Q_\tau/\average{Q_\tau}$,
one has for large $\tau$
\begin{equation}
f^{\mathtiny Q}_\tau(p_{\mathtiny Q})
= \left\{
   \begin{array}{ll}
     p_{\mathtiny Q} + O(\tau^{-1}) 
         &\mbox{if $0<p_{\mathtiny Q}<1$}\\
     p_{\mathtiny Q} -\frac14(p_{\mathtiny Q}-1)^2 + O(\tau^{-1})
         &\mbox{if $1<p_{\mathtiny Q}<3$}\\
     2 +O\left(\sqrt{(p_{\mathtiny Q}-3)/\tau}\right)
         &\mbox{if $p_{\mathtiny Q}>3$}.\\
\end{array}
\right.
\label{NFT}
\end{equation}
Here, we only gave the orders of magnitude of the finite-$\tau$
correction terms. Their detailed forms --- which differ in the
transient and the stationary state --- require an involved calculation
using the saddle-point method which can be found in
Ref.~\cite{VanZonCohen03b}. Note that these calculations need not be
redone for the current case but that we can make the substitutions in
Table~\ref{table} and consider the one-dimensional case as can be
obtained from footnote 24 of Ref.~\cite{VanZonCohen03b}.

Next, we discuss the \emph{work} fluctuations in this circuit. The total work
done in the circuit is the time integral of the power. The power is
the current through the circuit, $I$, times the voltage over the whole
circuit, $V_{cd}$, so that
\begin{eqnarray}
  W_\tau = \int_0^\tau dt\,I \left[ -\frac{q_t-It}{C} \right].
\label{workparalleleq}
\end{eqnarray}
This happens to be precisely the form of the work as we found in the
Brownian case\cite{VanZonCohen02b}, if we use Table~\ref{table}. This
is somewhat surprising because those analogues were based on
Eq.~\eq{eomparallel} which in principle only involves the current
through the resistor, while $W_\tau$ is the work done on the whole
circuit.  Since the work fluctuations for the Brownian particle
satisfy the conventional FT for
$\tau\to\infty$\cite{MazonkaJarzynski99,VanZonCohen02b}, by analogy we
know the same will hold here, i.e.,
\begin{eqnarray}
  f^{\mathtiny W}_\tau\equiv\frac{k_BT}{\average{W_\tau}} \ln \left[\frac{
P(+W_\tau)}{P(-W_\tau)}\right]
= \frac{p_{\mathtiny W}}{1-\varepsilon(\tau)},
\label{CFT}
\end{eqnarray}
where $p_{\mathtiny W}=W_\tau/\average{W_\tau}$. Here, for the TFT,
$\varepsilon=0$ while for the SSFT (for details see Ref.~\cite{VanZonCohen02b})
\begin{eqnarray}
  \varepsilon(\tau) = \frac{\tau_r(1-e^{-\tau/\tau_r})}{\tau},
\label{epsparallel}
\end{eqnarray}
where $\tau_r=RC$. Note that $\varepsilon(\tau)\to0$ for $\tau\to0$.
\begin{figure}[b]
\centerline{\includegraphics[height=0.21\textheight]{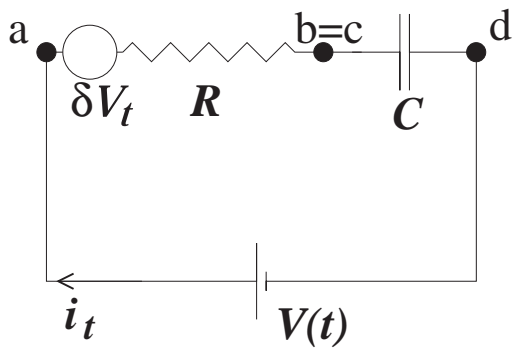}}
\caption{Serial RC circuit with Nyquist noise and imposed voltage.}
\label{seriesRCcircuit}
\end{figure}

We note that the behavior of the heat fluctuations in Eq.~\eq{NFT}
differs from that of the work fluctuations in Eq.~\eq{CFT} due to
exponential tails of the distribution of the heat
fluctuations\cite{VanZonCohen03a,VanZonCohen03b}, while those of
$P(W_\tau)$ are Gaussian.

A second example of a circuit that satisfies the extended heat
fluctuation theorem is depicted in Fig.~\ref{seriesRCcircuit}. Here,
the resistor with resistance $R$ and the capacitor with capacitance
$C$ are arranged \emph{in series} and are subject to a linearly
increasing voltage source $V(t)=At$. Again, there is a thermal noise
generator next to the resistance. The definitions of $i_t$ and $q_t$
are still that they are the current and charge through $R$ at time
$t$, respectively, but $q'_t=q_t$ here, since all charge that runs
through $R$ ends up in $C$.  We find the appropriate Langevin equation
as follows. The imposed voltage, $V(t)$, is equal to the
potential difference $V_{ad}= i_t R + \delta V_t + {q_t}/{C}$
(cf.~Fig.~\ref{seriesRCcircuit}).  Using $i_t=dq_t/dt$ and $V(t)=At$,
we find
\begin{eqnarray}
  0 = - R \ddh{q_t}{t} - \delta V_t - \frac{q_t-CA\,t}{C}  .
\label{eomseries}
\end{eqnarray}
This equation is the same as that of Eq.~\eq{eomparallel} except that
$I$ is replaced by $CA$.  This means that as far as the current
through the resistance is concerned, a constant current source with a
capacitor in parallel to it is equivalent to a linearly increasing
voltage with a capacitor in series with it.  It also means that this
case is analogous to the Brownian particle dragged through a fluid by
a harmonic potential as well.  Quantities for this circuit and the
Brownian model can therefore be translated into each other using table
\ref{table}, except for the work, as we will explain below.

Consider now the \emph{heat} developed in the resistor during a time $\tau$,
which is given by the time integral of the voltage over the resistor,
$V_{ab}$, times the current through it, $i_t$, so
\begin{eqnarray}
  Q_\tau = \int_0^\tau dt\,  i_t [i_t R + \delta V] =
-\int_0^\tau dt\,  i_t \frac{q_t-CAt}{C} ,
\end{eqnarray}
where we used Eq.~\eq{eomseries}.  Note that using $I\sim CA$, this
form for $Q_\tau$ is the same as in Eq.~\eq{5} for the parallel case.
Hence, the heat in the serial case behaves precisely the same as in
the parallel circuit, because we have seen that $q_t$ and $i_t$ behave
the same in both circuits.  Thus, we known that the heat in the serial
RC circuit, in the parallel RC circuit and in the Brownian system all
behave analogously, and all satisfy the extended
FT\cite{VanZonCohen03a,VanZonCohen03b} in Eq.~\eq{NFT}.

We will now consider the work fluctuations in the serial RC
circuit. The work is the time integral of the total current, $i_t$,
times the total voltage, $V(t)=At$, hence
\begin{eqnarray}
  W^{*}_\tau = \int_0^\tau dt\, i_t A t.
\label{workserieseq}
\end{eqnarray}
Even when we use $I\sim CA$, this is not the same form as $W_\tau$ in
Eq.~\eq{workparalleleq} (which is why we added a superscript $^*$),
and likewise, using table \ref{table}, it is not of the same form as
in the Brownian case. Clearly, we cannot use the same results for
$W^*_\tau$ as we obtained for $W_\tau$ from the Brownian case or the
parallel circuit: we in fact need a additional calculation.  For this,
we use the method in Ref.~\cite{VanZonCohen02b}. For a Gaussian
$P(W_\tau^*)$ one has
\begin{eqnarray}
  f^{\mathtiny W*}_\tau\equiv\frac{k_BT}{\average{W^*_\tau}} \ln \left[\frac{
P(+W^*_\tau)}{P(-W^*_\tau)}\right]
= \frac{p_{\mathtiny W}^*}{1-\varepsilon(\tau)},
\end{eqnarray}
where $p_{\mathtiny W}^*=W^*_\tau/\average{W^*_\tau}$ and
$\varepsilon(\tau)=1-V/(2M)$ with $M=\average{W_\tau}$ and
$V=\average{(W_\tau-\average{W_\tau})^2}$.  The quantities $M$ and $V$
in turn are calculated using the definition of $W^*_\tau$ in
Eq.~\eq{workserieseq} and the relations $\average{q_t} =
CA[t-\tau_r(1-e^{-t/\tau_r})]$ and
$\average{(q_t-\average{q_t})(q_{t'}-\average{q_{t'}})} =
k_BTC\,e^{-|t-t'|/\tau_r}$ (from Ref.~\cite{VanZonCohen02b}, using
Table~\ref{table}). This yields
\begin{eqnarray}
  \varepsilon(\tau) = 2 \frac{\tau_r^2(1-e^{-\tau/\tau_r})-\tau\tau_r
  e^{-\tau/\tau_r}}{\tau^2}.
\label{epsseries}
\end{eqnarray}
Note that this goes to zero asymptotically as $1/\tau^2$, which is
faster than the $1/\tau$ decay of $\varepsilon(\tau)$ in
Eq.~\eq{epsparallel}.

These result so far made no explicit reference to nanoscale
circuits. Indeed, these results are valid for a circuit of any
size. We will now turn to their relevance for nanocircuits.  The
extended FT shows that large heat fluctuations are more likely to
occur than according to the conventional FT, due to the exponential
tails of the distribution of heat fluctuations. This could be
important for the design of nanostructures because of the rise of
temperature due to heat development. Either the \emph{average} current
through $R$ could be too large or a \emph{large heat fluctuation}
might occur.  Fluctuations of the energy exist already in an
equilibrium system in contact with a heat reservoir, i.e. a circuit
for which $I=0$. In that case, the energy fluctuations of an $N$
atomic resistor will be of the order of $\sqrt{N} k_BT$. To explore
the properties of nanostructures, composed of just a few atoms, we
start with $N=1$. For that case, energy fluctuations are of the order
of $k_BT$ which, if they were not removed, could amount to a
significant increase or decrease of the resistor's temperature. An
estimate of this temperature change can be made using the law of
Dulong and Petit that (at room temperatures) the specific heat per
atom of a solid is $3k_B$. Thus, the change in temperature could be as
large as $\Delta T = k_BT/(3k_B) = T/3= 100 \mathrm K$, for $T= 300
\mathrm K$.

In our Langevin theory for the circuits, we have assumed a constant
temperature. To still be able to use the theory, we need that the
temperature does not vary too much. A similar condition is needed if
the resistor is not to fail. To satisfy this condition, the heat
developed will have be transported away at a fast enough time scale
$\tau_{\mathtiny T}$. If the heat $Q_{\tau_{\mathtiny T}}$ developed
in this time is large enough to significantly increase the material's
temperature, failure may occur. The induced temperature difference,
given by $\Delta T=Q_{\tau_{\mathtiny T}}/C_V$, is insignificant if
$\Delta T/T\ll 1$, i.e., if
\begin{equation}
  \frac{Q_{\tau_{\mathtiny T}}}{C_VT} \ll 1.
\label{1}
\end{equation}
This condition should hold both for the average and the
fluctuations of $Q_{\tau_{\mathtiny T}}$.  Let us first consider the
average (i.e., what one would do for macroscopic circuits). In the
stationary state of the parallel circuit, the average heat is
$Q_{\tau_{\mathtiny T}}=IV\tau_{\mathtiny T}$, where $V=IR$, so
\begin{equation}
  \frac{Q_{\tau_{\mathtiny T}}}{C_VT} = \frac{I^2 R\tau_{\mathtiny T}}{C_V T} =
  \frac{\tau_{\mathtiny T}}{\tau_{\mathtiny H}},
\label{2}
\end{equation}
where $\tau_{\mathtiny H} = {C_V T}/{I^2 R}$. (This in fact also holds
for the serial circuit if we replace $I$ by $CA$.)  Equation
(\ref{1}) now becomes
\begin{equation}
  \tau_{\mathtiny T} \ll \tau_{\mathtiny H}.
\label{4}
\end{equation}
That is, the time scale of temperature relaxation given by
$\tau_{\mathtiny T}$ has to be fast compared to the heating time
$\tau_{\mathtiny H}$.  However, even if the average of the heat is
well-behaved, the heat fluctuations might still damage the circuit. We
need the typical size of these fluctuations.  From our
previous work on the FTs, we found that for long times $\tau$,
$\sqrt{\average{(Q_{\tau}-\average{Q_\tau})^2}}\approx \sqrt{2
k_BT\average{Q_\tau}}$\cite{VanZonCohen03a,VanZonCohen03b}.  Thus, for
$\tau=\tau_{\mathtiny T}$, $\sqrt{2 k_B T I^2R\tau_{\mathtiny T}}$ is
a typical value for $Q_{\tau_{\mathtiny T}}$ which we can insert into
Eq.~(\ref{1}), giving
\begin{equation}
\label{5p}
  \frac{2k_BI^2R\tau_{\mathtiny T}}{TC_V^2} \equiv \frac{\tau_{\mathtiny T}}{\tau_{\mathtiny F}}\ll 1
\end{equation}
where the time scale for the heat fluctuations is $ \tau_{\mathtiny F}
= {TC_V^2}/{2k_B I^2R} = ({C_V}/{2k_B}) \tau_{\mathtiny H}$.  For an
$N$ atomic solid at room temperature, for which Dulong-Petit is valid,
we know that $\tau_{\mathtiny F}/\tau_{\mathtiny H} = C_V/2k_B$ =
$\frac32 N$. Thus, the factor between $\tau_{\mathtiny F}$ and
$\tau_{\mathtiny H}$ is always bigger than one, so that Eq.~(\ref{5p})
in fact follows from Eq.~(\ref{4}).  As a result, if on average the circuit
will not fail, then the fluctuations will not make it fail either.
However, for systems below room temperature, for which quantum
effects become relevant, the ratio $C_V/k_B$ can be much smaller than
one (i.e. Debye's law), opening up the possibility that in
that case the requirement on $\tau_{\mathtiny F}$ in Eq.~\eq{5p} could
be stricter than that on $\tau_{\mathtiny H}$ in Eq.~\eq{4}.

Although reassuring, Eq.~\eq{5p} concerns only typical
fluctuations. However, one also needs to be concerned with large
fluctuations. Compared to the Gaussian distributed work (or power)
fluctuations, large fluctuations for heat are much more likely due to
the exponential tails of its distribution function. Furthermore, if
the condition in Eq.~\eq{1} is not met, temperature variations will
occur and the theory then needs to include a coupling to the heat
diffusion equation. This could be relevant for future experiments.

In conclusion, we studied work and heat fluctuations in electric
circuits using analogies to Brownian systems with non-universal
additions to the Nyquist noise-Brownian motion analogy.  Our analogy
links the work and heat fluctuations in a parallel RC circuit to those
in a Brownian system for which the work and heat fluctuations are
known to satisfy the conventional FT and extended FT respectively.
For the serial circuit, the analogy also works for the heat
fluctuations, but not for the work fluctuations. However, a short
calculation [below Eq.~\eq{workserieseq}] shows they still satisfy the
conventional FT.

RVZ and EGDC acknowledge the support of the Office of Basic
Engineering of the US Department of Energy, under grant No.\
DE-FG-02-88-ER13847.


\begin{thebibliography}{99}
\bibitem{footnote11}
 
A self-inductance $L$ in series with $R$ can result in a mass-like
term in Eqs.~\eq{eomparallel} and \eq{eomseries}.  To know the status
of the FTs for $L\neq 0$ would require an additional calculation.
\end{thebibliography}
\end{document}